\documentclass[aps,showpacs,twocolumn,superscriptaddress]{revtex4}

\usepackage{graphicx}
\usepackage{dcolumn}

\begin{document}

\title{Comment on ``Five-Body Cluster Structure of the Double-$\Lambda$ 
Hypernucleus $_{\Lambda\Lambda}^{~11}{\rm Be}$"} 

\author{A.~Gal} 
\affiliation{Racah Institute of Physics, The Hebrew University, 
Jerusalem 91904, Israel}  
\affiliation{ECT$^{\ast}$, Villa Tambosi, I-38100 Villazzano (Trento), 
Italy}

\author{D.J.~Millener}
\affiliation{Physics Department, Brookhaven National Laboratory, Upton, 
NY 11973, USA} 

\begin{abstract}
\rule{0ex}{3ex} 

Hiyama {\it et al.} \cite{hiyama10} have recently reported on a pioneering 
five-body $\alpha\alpha n \Lambda\Lambda$ cluster-model (CM) calculation 
of $_{\Lambda\Lambda}^{~11}{\rm Be}$ in order to confront a possible 
interpretation of the KEK-E373 HIDA event \cite{nakazawa10}. 
Unfortunately, a six-body $\alpha\alpha nn \Lambda\Lambda$ calculation of 
$_{\Lambda\Lambda}^{~12}{\rm Be}$ to confront another possible 
interpretation is beyond reach at present. Using experimental $B_{\Lambda}$ 
values with small corrections based on recently determined $\Lambda N$ 
spin-dependent interaction parameters \cite{millener10}, 
we obtain binding-energy shell-model (SM) estimates for both 
$_{~~\Lambda\Lambda}^{11,12}{\rm Be}$, concluding that neither 
$_{\Lambda\Lambda}^{~11}{\rm Be}$ nor $_{\Lambda\Lambda}^{~12}{\rm Be}$ 
provide satisfactory interpretation of the HIDA event. 
The SM approach is tested by reproducing 
$B_{\Lambda\Lambda}^{\rm exp}({_{\Lambda\Lambda}^{~13}{\rm B}})$. 

\end{abstract} 

\pacs{21.80.+a, 21.60.Cs, 21.60.Gx} 

\maketitle 

The input to the SM estimates consists of three $\Lambda$-spin-dependent 
$\Lambda N$ interaction parameters ($\Delta, S_{\Lambda}, T$) fitted 
to the six known $\Lambda$ hypernuclear doublet splittings beyond 
$_{\Lambda}^{9}{\rm Be}$ and of the induced nuclear spin-orbit 
parameter $S_{N}$ extracted from the excitation energy of 
$_{~\Lambda}^{16}{\rm O}(1_{2}^{-})$. The fit also includes a $\Lambda-\Sigma$ 
coupling interaction \cite{millener10}. For this fit, with a spin-independent 
$\Lambda N$ interaction parameter ${\overline V}_{\Lambda N} = -1.04$ MeV, 
ground-state (g.s.) binding energies of $\Lambda$ hypernuclei with mass number 
$A=10,11,12$ are reproduced to within $\delta B_{\Lambda}^{\rm SM}\lesssim 
0.2$~MeV. The associated SM estimate for the $\Lambda\Lambda$ binding energy 
of the $\Lambda\Lambda$ hypernucleus $_{\Lambda\Lambda}^{~~\rm A}{\rm Z}$ is 
given by 
\begin{equation} 
B_{\Lambda\Lambda}^{\rm SM}({_{\Lambda\Lambda}^{~~\rm A}{\rm Z}}) = 
2{\overline B}_{\Lambda}^{\rm SM}({_{~~~\Lambda}^{\rm A-1}{\rm Z}}) + 
{\langle V_{\Lambda\Lambda} \rangle}_{\rm SM}, 
\label{eq:BLL} 
\end{equation} 
where ${\overline B}_{\Lambda}^{\rm SM}({_{~~~\Lambda}^{\rm A-1}{\rm Z}})$ is 
the $(2J+1)$-averaged binding energy of the g.s. doublet in the $\Lambda$ 
hypernucleus $_{~~~\Lambda}^{\rm A-1}{\rm Z}$, as appropriate to the spin zero 
$(1s_{\Lambda})^2$ configuration of $_{\Lambda\Lambda}^{~~\rm A}{\rm Z}$. 
The $\Lambda\Lambda$ interaction contribution to 
$B_{\Lambda\Lambda}({_{\Lambda\Lambda}^{~~\rm A}{\rm Z}})$ is deduced from 
the NAGARA event \cite{nakazawa10}: 
${\langle V_{\Lambda\Lambda}\rangle}_{\rm SM} = 
B_{\Lambda\Lambda}({_{\Lambda\Lambda}^{~~6}{\rm He}}) - 
2B_{\Lambda}({_{\Lambda}^{5}{\rm He}}) = (0.67\pm 0.17)$ MeV, 
close to $\langle V_{\Lambda\Lambda}^{\rm CM}\rangle = B_{\Lambda\Lambda}
(V^{\rm CM}_{\Lambda\Lambda})-B_{\Lambda\Lambda}(V_{\Lambda\Lambda}=0)
\approx 0.55$ MeV, with $V^{\rm CM}_{\Lambda\Lambda}$ also fitted to 
$B_{\Lambda\Lambda}({_{\Lambda\Lambda}^{~~6}{\rm He}})$ \cite{hiyama10}. 
Table~\ref{tab:summary} lists 
${\overline B}_{\Lambda}^{\rm SM}({_{~~~\Lambda}^{\rm A-1}{\rm Z}})$ 
input to Eq.~(\ref{eq:BLL}), constrained by $B_{\Lambda}^{\rm exp}
({_{~~~\Lambda}^{\rm A-1}{\rm Z}})$ values \cite{davis05}, 
plus $B_{\Lambda\Lambda}^{\rm SM}({_{\Lambda\Lambda}^{~\rm A}{\rm Z}})$ 
predictions. 

\begin{table} [h] 
\caption{SM input and 
$B_{\Lambda\Lambda}^{\rm SM}({_{\Lambda\Lambda}^{~~\rm A}{\rm Z}})$ 
predictions (in MeV).} 
\label{tab:summary} 
\begin{tabular}{cccc} 
\hline \hline 
~$_{\Lambda\Lambda}^{~~\rm A}{\rm Z}$~ & 
~${\overline B}_{\Lambda}^{\rm SM}({_{~~~\Lambda}^{\rm A-1}{\rm Z}})$~ & 
~$B_{\Lambda\Lambda}^{\rm SM}({_{\Lambda\Lambda}^{~~\rm A}{\rm Z}})$~ & 
~$B_{\Lambda\Lambda}^{\rm exp}({_{\Lambda\Lambda}^{~~\rm A}{\rm Z}})$~ 
\cite{nakazawa10}\\
\hline 
~$_{\Lambda\Lambda}^{~11}{\rm Be}$~ & ~$8.86\pm 0.10$~ & ~$18.39\pm 0.20$~ & 
~$20.83\pm 1.27$~ \\ 
~$_{\Lambda\Lambda}^{~12}{\rm Be}$~ & ~$10.02\pm 0.05$~ & ~$20.71\pm 0.20$~ & 
~$22.48\pm 1.21$~ \\ 
~$_{\Lambda\Lambda}^{~13}{\rm B}$~ & ~$11.27\pm 0.06$~ & ~$23.21\pm 0.21$~ & 
~$23.3\pm 0.7$~ \\ 
\hline \hline 
\end{tabular} 
\end{table} 

For the calculation of 
$B_{\Lambda\Lambda}^{\rm SM}({_{\Lambda\Lambda}^{~11}{\rm Be}})$, 
since our SM fit maintains charge symmetry, we averaged statistically on 
$B_{\Lambda}^{\rm exp}({_{~\Lambda}^{10}{\rm Be}_{\rm g.s.}})$ and 
$B_{\Lambda}^{\rm exp}({_{~\Lambda}^{10}{\rm B}_{\rm g.s.}})$ \cite{davis05} 
to get a SM input value $B_{\Lambda}^{\rm SM}({_{~\Lambda}^{10}{\rm Be}})=
(8.94\pm 0.10)$ MeV. The SM prediction in Table~\ref{tab:summary} compares 
well with the CM prediction 
$B_{\Lambda\Lambda}^{\rm CM}({_{\Lambda\Lambda}^{~11}{\rm Be}})=18.23$ MeV 
\cite{hiyama10} in spite of the differing input. However, a meaningful 
comparison requires using identical interactions. For example, the induced 
nuclear spin-orbit interaction (parameter $S_{N}$), known to play a key role 
in $p$ shell $\Lambda$ hypernuclei \cite{millener10}, contributes close to 
400 keV to $B_{\Lambda}^{\rm SM}({_{~\Lambda}^{10}{\rm Be}_{\rm g.s.}})$ 
and twice as much to 
$B_{\Lambda\Lambda}^{\rm SM}({_{\Lambda\Lambda}^{~11}{\rm Be}})$, 
but it is missing in the CM works \cite{hiyama10,hiyama02}. 

For the calculation of 
$B_{\Lambda\Lambda}^{\rm SM}({_{\Lambda\Lambda}^{~12}{\rm Be}})$, we replaced 
the spin dependent and $\Lambda-\Sigma$ coupling contributions to 
$B_{\Lambda}^{\rm exp}({_{~\Lambda}^{11}{\rm B}_{\rm g.s.}})$ \cite{davis05} 
by those appropriate to $_{~\Lambda}^{11}{\rm Be}_{\rm g.s.}$. 
For the calculation of 
$B_{\Lambda\Lambda}^{\rm SM}({_{\Lambda\Lambda}^{~13}{\rm B}})$, since the 
value of $B_{\Lambda}^{\rm exp}({_{~\Lambda}^{12}{\rm C}_{\rm g.s.}})$ is 
controversial, we used 
$B_{\Lambda}^{\rm exp}({_{~\Lambda}^{12}{\rm B}_{\rm g.s.}})$ \cite{davis05} 
plus a 161 keV ($1^{-}_{\rm g.s.},2^{-}_{\rm exc}$) doublet splitting from 
$_{~\Lambda}^{12}{\rm C}$ \cite{tamura10}. 

The excellent agreement between 
$B_{\Lambda\Lambda}^{\rm SM}({_{\Lambda\Lambda}^{~13}{\rm B}})$ and 
$B_{\Lambda\Lambda}^{\rm exp}({_{\Lambda\Lambda}^{~13}{\rm B}})$ provides 
a consistency check on the SM estimates 
$B_{\Lambda\Lambda}^{\rm SM}({_{~~\Lambda\Lambda}^{11,12}{\rm Be}})$ 
listed in Table~\ref{tab:summary}. Comparing these estimates with the 
corresponding $B_{\Lambda\Lambda}^{\rm exp}$ options listed in the table, 
we conclude that a $_{\Lambda\Lambda}^{~12}{\rm Be}$ assignment to the HIDA 
event is no more likely than a $_{\Lambda\Lambda}^{~11}{\rm Be}$ assignment. 

Useful discussions with Emiko Hiyama are gratefully acknowledged. AG thanks 
ECT$^{\ast}$ Director Achim Richter for hospitality when this Comment was 
conceived. D.J.M. acknowledges the support by the U.S. DOE under Contract 
DE-AC02-98CH10886 with the Brookhaven National Laboratory.

\end{document}